# Graphene or h-BN paraffin composite structures for the thermal management of Li-ion batteries: A multiscale investigation

Bohayra Mortazavi[*,1], Hongliu Yang[2], Farzad Mohebbi[1], Gianaurelio Cuniberti[2], Timon Rabczuk[3,#]

[1]*Institute of Structural Mechanics, Bauhaus-Universität Weimar, Marienstr. 15, D-99423 Weimar, Germany.*
[2]*Institute for Materials Science and Max Bergman Center of Biomaterials, TU Dresden, 01062 Dresden, Germany.*
[3]*College of Civil Engineering, Department of Geotechnical Engineering, Tongji University, Shanghai, China*

**Abstract**

The reliability and safety of lithium-ion batteries can be affected by overheating issues. Phase change materials like paraffin due to their large heat capacities are among the best solutions for the thermal management of batteries. In this investigation, multiscale modelling techniques were developed to explore the efficiency in the thermal management of rechargeable batteries through employing the paraffin composite structures. A combined atomistic-continuum multiscale modelling was conducted to evaluate the thermal conductivity of paraffin reinforced with graphene or hexagonal boron-nitride nanosheet additives. In addition, heat generation during a battery service was simulated using the Newman's electrochemical model. Finally, three-dimensional heat transfer models were constructed to investigate the effectiveness of various paraffin composite structures in the thermal management of a battery system. Interestingly, it was found that the thermal conductivity of paraffin nanocomposites can be enhanced by several times but that does not yield significant improvement in the batteries thermal management over the pure paraffin. The acquired findings can be useful not only for the modelling of nanocomposites but more importantly for the improvement of phase change materials design to enhance the thermal management of rechargeable batteries and other electronic devices.

*Corresponding author (Bohayra Mortazavi): bohayra.mortazavi@gmail.com
Tel: +49 157 8037 8770;
Fax: +49 364 358 4511
#Timon.rabczuk@uni-weimar.de

## 1. Introduction

In recent years, development of Lithium-ion (Li-ion) batteries technology are considered as crucial components of progress in electronic and communication devices and automobile industries. When using Li-ion batteries, however an adverse condition termed overheating may happen which consequently declines the cell performance and may cause severe problems such as fire, explosion, and thermal runaway which, can lead to serious health and safety problems to the community [1,2]. In Li-ion battery packs, the need for dissipating the extra heat generated beyond the normal range deserves much attention to the thermal management. The use of phase change materials (PCM) in thermal management of Li-ion battery packs is becoming widespread due to their high heat capacity and capability to absorb and release heat in melting/solidification processes. The use of PCMs in a Li-ion battery pack delays the temperature rise inside the batteries, thereby decreasing the risk of overheating and thermal runaway as well [3]. Among the drawbacks of commonly used PCMs is their low thermal conductivity, motivating the use of composite structures through addition of highly thermal conductive fillers inside the PCMs [4–7].

To enhance the thermal conduction properties of PCM, fabrication of nanocomposite structures are among the most appealing approaches. In this case, nanosized fillers with remarkably high thermal conduction properties are distributed inside the polymer matrix to enhance the thermal conductivity of the mixture [8–10]. During the last decade, graphene [11,12] the planar form of $sp^2$ carbon atoms has attracted tremendous attentions owing to its exceptional mechanical, electrical and heat conduction properties. Incorporation of graphene inside the polymer matrix therefore enhances simultaneously the thermal conductivity, mechanical rigidity, and electrical properties [9,13–15]. The polymer nanocomposites made from graphene fillers are also probably electrical conductors which may cause undesirable effects in electronic devices. Hexagonal boron-nitride (h-BN) [16] so called "white graphene" is another atomic thick material that is electrically insulator with a wide bandgap. Among the experimentally fabricated nanostructured materials, h-BN nanomembranes are reported to present remarkably high thermal conductivities in a range of 250-600 W/mK [17,18]. Insulating property of h-BN films is also a desirable characteristic and such that their polymer nanocomposites are electrically an insulator.

As a matter of fact, the fabrication of composite structures can improve the thermal conductivity of PCM. In this regard, several experimental studies have been already conducted for the investigation of the thermal properties of paraffin reinforced with the graphene nanomembranes [19–21]. However, the efficiency in the improvement of the thermal management of Li-ion batteries through employing the PCM composite structures with different filler types, concentrations and microstructures is yet necessary to be explored more in detail. To address this issue, in the present investigation hierarchical multiscale modelling were conducted. In this scenario, nanostructural effects were taken into consideration on the thermal transport of PCM nanocomposites. We note that thermal conductivity of graphene/paraffin nanocomposites has been already simulated using the classical molecular dynamics (MD) simulations [22,23]. However, because of the computational costs of MD simulations, the sizes of the studied samples have been far from reality. To more realistically evaluate the thermal conductivity of paraffin matrix filled with graphene or h-BN nanofilms, we developed a multiscale modelling method. To this aim, MD simulations were conducted to evaluate the thermal contact conductance (Kapitza conductance) between the fillers and the paraffin matrix. Based on the MD results, nanocomposites representative volume elements (RVE) were then constructed using the finite element (FE) modelling to simulate the effective thermal conductivity. Multiscale modelling results highlight several critical aspects in the modelling of effective thermal conductivity of nanocomposites. Interestingly, h-BN paraffin nanocomposites are discussed that may present higher thermal conductivity as well as heat capacity in comparison with the nanocomposites filled with the graphene particles.

Next, the temperature rise in a battery system was explored during various charging or discharging rates. In this case, the electrochemical reactions and subsequent heat generations during charging/discharging cycles of a Li-ion battery were simulated using the Newman's electrochemical pseudo two-dimensional (2D) model [24]. Finally, the efficiency in the thermal management of a battery system was discussed through employing no PCM, pure paraffin, paraffin nanocomposites and graphite network paraffin composite. The multiscale modelling results could provide useful insights with respect to the thermal management of not only the Li-ion batteries but also other systems such as nanoelectronics through employing the polymer nanocomposites.

In the present investigation, first the developed modelling techniques and their preliminary results are presented and then the effects of paraffin composite structures on the thermal management of a battery system will be discussed.

## 2. Multiscale modelling and results

### 2.1 Methodology

The multiscale modelling scheme conducted in this study is illustrated in Fig. 1. To simulate the electrochemical processes of a Li-ion battery at different charging or discharging loadings, a model within the framework of Newman's pseudo 2D electrochemical model [24] was developed. As shown in Fig. 1a, in this model to simulate the battery operation, its three-dimensional (3D) structure is represented using a one-dimensional (1D) model. In the next step, reversible and irreversible volumetric heat generations during the battery charging or discharging process were acquired according to the electrochemical response. Such that the main goal from the electrochemical modelling was to obtain total average volumetric heat generation ($Q_{avg}$ as illustrated in Fig. 1a) for a battery under various loading conditions. These volumetric heat generation functions were then finally used in our 3D finite element heat transfer model to investigate the temperature rise in a battery system.

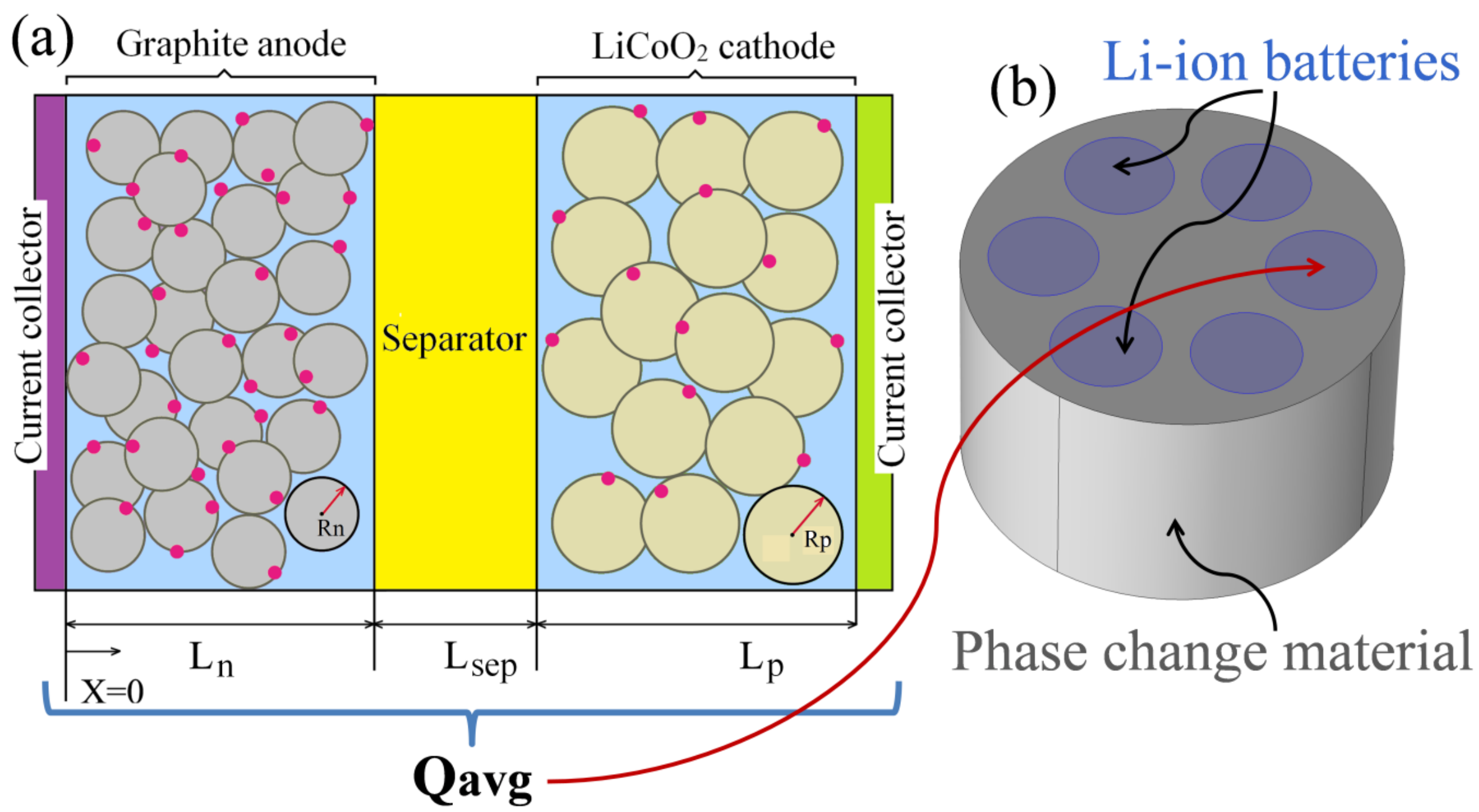

Fig.1, Schematic illustration of multiscale modelling developed in this work. (a) Newman's electrochemical model was used to acquire the total average volumetric heat generation ($Q_{avg}$) during the battery service. In this model, the battery 3D components (anode, separator and cathode) are represented using the 1D models. (b) Developed 3D finite element heat transfer model to explore the temperature rise in a battery system with six Li-ion batteries embedded in the phase change material (PCM).

As depicted in Fig. 1b, in our modelling of a battery system, 6 Li-ion batteries embedded in the PCM were simulated. The diameter and height of each Li-ion battery were assumed to be 40 mm and 92 mm, respectively, based on the Hitachi's batteries for hybrid electric vehicles (with capacity of 4.4 Ah) [25]. The thermal conductivity, heat capacity, and density of each Li-ion battery were considered to be 3.4 W/mK, 1.280 J/gK and 2680 kg/$m^3$, respectively [26]. We note that more accurate modelling can be achieved by considering complete Li-ion batteries components. However, since the objective of present study is to compare different paraffin composites structures in the thermal management, the main findings are convincingly independent of the details in the modelling of batteries structure.

In this study, the PCM was considered to be pure paraffin or paraffin reinforced with graphene or h-BN nanomembranes or graphite networks and rings. The diameter of PCM was then considered to be 160 mm with the same height as that of the Li-ion batteries. In order to obtain the thermal conductivity of paraffin nanocomposites, combined molecular dynamics and finite element multiscale [13] modelling were performed. The heat capacity of the nanocomposites was also evaluated by employing the rule of mixtures.

In this section, first the electrochemical modelling and the corresponding results are explained. Then, the multiscale modelling for the thermal properties of paraffin nanocomposites are presented and the main findings are discussed.

### 2.2 Mathematical modelling and results for the heat generation in a Li-ion battery

#### 2.2.1 Electrochemical modelling

As shown in Fig. 1a, a common Li-ion battery consists of two current collectors, the positive electrode, the electron-blocking separator, and the negative electrode. The negative and positive electrodes and separator are all porous components which are filled by a lithiated organic solution that serves as the electrolyte. Based on the pseudo 2D model of Newman, the dynamical performance of a Li-ion cell is characterized by solving four coupled partial differential equations (PDE) (see Table 1). These equations describe the time evolution of the lithium (Li) concentration profile and potentials inside the electrode and electrolyte phases, under charge conservation. In this scheme, in the modelling of a Li-ion battery, three PDEs are considered as 1D problems which are coupled to another 2D model to simulate the Li atoms diffusion inside the solid particles. In the present investigation, the negative and positive electrodes were assumed to be the meso-carbon micro-bead (MCMB) graphite and $LiCoO_2$, respectively [27]. Since the Newman's pseudo 2D electrochemical model has been widely used for the modeling of Li-ion batteries performance [27–29], the governing differential equations and the corresponding boundary conditions are briefly summarized in Table 1. The effective properties were simply obtained using the Bruggeman's relation [30] which are also summarized in Table 1.

Bulter-Volmer equation was used to simulate the charge transfer rate at the interface between the active particles and the electrolyte as follows [24]:

$$j^{Li} = a_s i_0 \left\{ \exp\left[\frac{0.5F}{RT}\eta\right] - \exp\left[-\frac{0.5F}{RT}\eta\right] \right\} \qquad (1)$$

where $a_s$ is the active surface area per electrode unit volume and the $i_0$ is the exchange current density which is introduced as follows [24]:

$$i_0 = K_i (C_e)^{0.5} (C_{s,\max} - C_{s,surf})^{0.5} (C_{s,surf})^{0.5} \qquad (2)$$

where $K_i$ is the Reaction rate coefficient, $C_{s,max}$ is the maximum Li concentration in the solid phase particles, $C_{s,surf}$ is the concentration of Li at the surface of solid phase particles, and $C_e$ is the Li-ion concentration in electrolyte. The over-potential $\eta$ in Eq. 1 is given by [24]:

$$\eta = \phi_s - \phi_e - U \qquad (3)$$

where $U$ is the equilibrium potential which is a function of intercalated Li and is on the basis of empirical functions. In this work, the equilibrium potentials for the negative and positive electrodes and the electrolyte's ionic conductivity and salt diffusion coefficient as a function of Li concentration at the room temperature were extracted from the work by Kumaresan *et al.* [27]. The parameter set used in our electrochemical modeling is listed in Table 2. It is worthy to mention that, based on the considered Li-ion battery characteristics in Table 2, the current density for the 1C current was calculated to be around 24 $A/m^2$. In order to solve the aforementioned mathematical equations in COMSOL Multiphysics, a 1D model with 3 line segments representing the negative electrode, the separator, and the positive electrode was developed. In this regard, the potential in the solid phase and electrolyte, the mass balance relationship for the electrolyte and reaction current density are solved using the 1D model. To simulate the diffusion inside the solid particles, two rectangles were considered to depict the Li concentration profile inside the electrode's active material as a function of particle's average radius. The reaction current density obtained from the 1D solution was projected to the top boundary of the 2D geometry by using “subdomain extrusion coupling variables” [29]. Our implementation of Newman's model in COMSOL was validated in the earlier study [31], by comparing with the experimental and theoretical data by Doyle *et al.* [32]. We note that for the simplicity of the Li-ion battery modeling, simulations were achieved by assuming a constant temperature (T = 298 K) and such that the effects of temperature changes on the electrochemical response of the Li-ion battery were neglected.

### 2.2.2 Heat generation

Typically, the heat generation in Li-ion batteries can be attributed to three main sources: the heat from the reaction current and over-potentials ($Q_r$), the ionic ohmic heat due to the current carried in solid and electrolyte phases ($Q_j$), and the reversible heat from entropy changes in solid particles ($Q_{rev}$). The irreversible heat generated during the battery operation is therefore the summation of reaction and ohmic heats. The total volumetric heat sources were calculated by integrating the local volume-specific reaction heat across the 1D cell domain and multiplying by cell area, *A*. In this study, heat sources are introduced as follows [33]:

$$Q_r = Aj^{Li}(\phi_s - \phi_e - U) \quad (4)$$

$$Q_j = A\sigma^{eff}\left(\frac{\partial \phi_s}{\partial x}\right)^2 + A\kappa^{eff}\left(\frac{\partial \phi_e}{\partial x}\right)^2 + A\kappa_D^{eff}\frac{\partial \ln C_e}{\partial x}\left(\frac{\partial \phi_e}{\partial x}\right) \quad (5)$$

$$Q_{rev} = j^{Li}T\frac{\partial U}{\partial T} \quad (6)$$

In Eq. 5, the first term expresses the ohmic heat of the solid phase; the second and third terms express the heat generation inside the electrolyte phase. Eq. 6 presents the reversible heat generation originated from the entropy change ($\partial U/\partial T$) in active electrode materials which can be either positive or negative. In the present work, $\partial U/\partial T$ curves as a function of Li concentration inside graphite and $LiCoO_2$ active solid particles were adopted from the work by Kumaresan *et al*. [27]. The estimated heat generation rates based on the Newman's model were then used to simulate the temperature rise in a battery system with six individual batteries as shown in Fig. 1b. To simulate the thermal transfer to the ambient, convective heat transfer boundary condition were introduced over the outer surfaces of the constructed 3D heat transfer model (see Fig. 1b). In this case different convective heat transfer coefficients were assumed.

**2.2.3 Li-ion battery modelling results**

The simulated cell voltages as a function of time for various charging or discharging current densities with respect to the 1C current are illustrated in Fig. 2. Here, the cell voltage was obtained using: $V = \phi_s\big|_{x=\mathrm{Ln+Lsep+Lp}} - \phi_s\big|_{x=0}$ . In these simulations, the cell was discharged or charged starting from the almost fully charged or discharged states, respectively. In our modeling, for the 3C current the simulation was terminated as the voltage goes further than 4.8 V or below 3.3 V for charging and discharging, respectively. As expected, for the both charging and discharging by increasing the current density the initial jump in the voltage increases which means that the over-potential to reach the desired current is increased. Because of different reaction rate coefficients for cathode and anode materials, this initial jump in the voltage profile is clearly higher for charging as compared to discharging cycle.

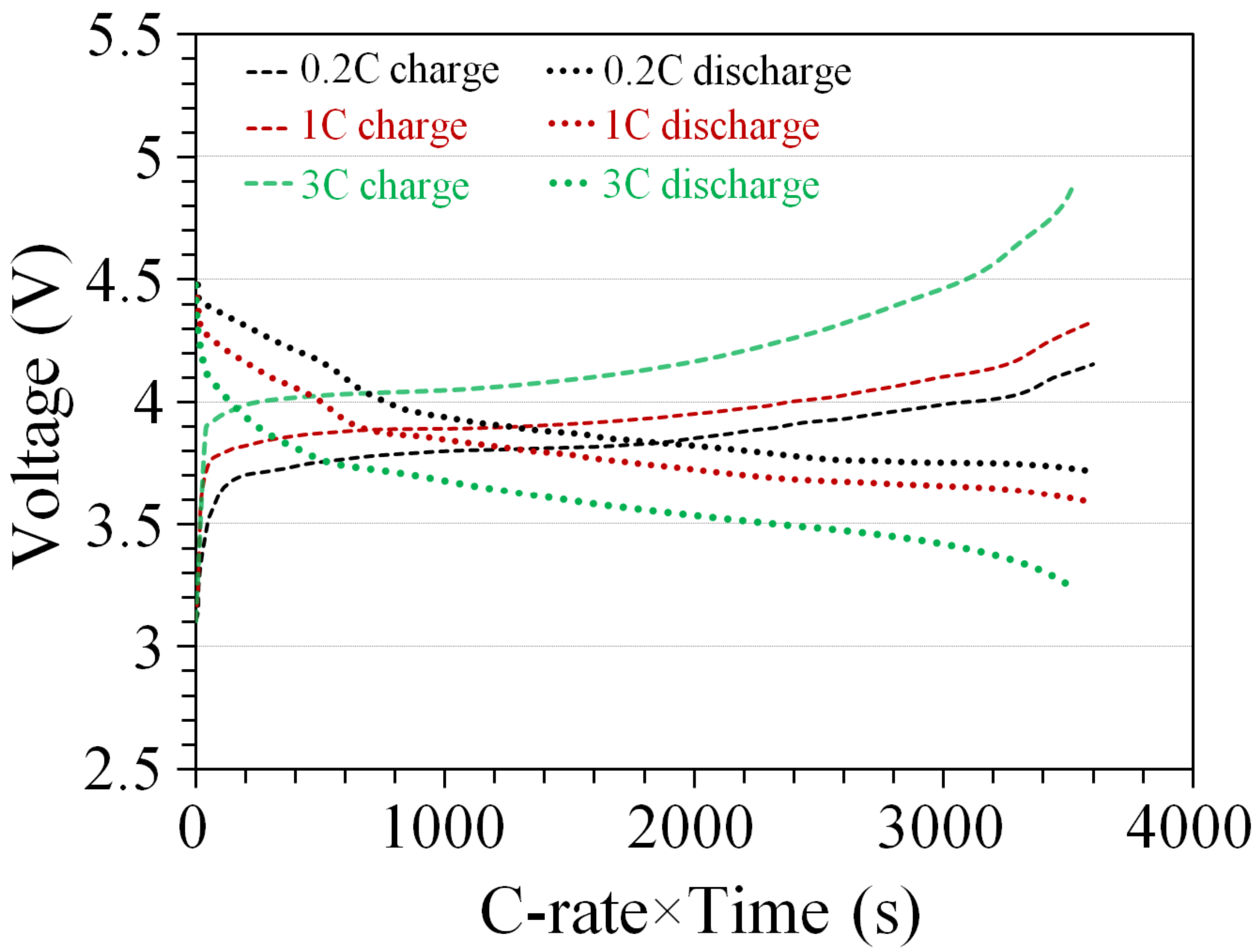

Fig.2, Simulated cell voltages as a function of time for various charging or discharging current densities with respect to the 1C current (24 A/$m^2$).

In this work, the volumetric heat generations for a Li-ion battery were calculated for 0.2C, 1C and 3C current densities for both charging and discharging cycles. The calculated total average volumetric reversible and irreversible heat generation for a 1C charging and discharging current as a function of cycle time are illustrated in Fig. 3. The heat generated during the battery operation can be explained by analyzing the heat generation mechanism in the energy balance equation. In Eq. 6, the reversible heat $IT\left(\frac{\partial U_p}{\partial T}-\frac{\partial U_n}{\partial T}\right)$ could be either positive or negative depending on the charging or discharging process and the state of the charge, whereas the irreversible heat is always positive independent of the current direction. As it is obvious, for the reversible heat the magnitude of $\partial U/\partial T$ directly defines the magnitude of the heat generation. We remind that the experimental results [27,34] for the both cathode and anode material reveal non-uniform patterns for the $\partial U/\partial T$ curves and as the state of the charge changes they take positive or negative values. Compared to graphite, however, the magnitude of $\partial U/\partial T$ curves for $LiCoO_2$ are larger [27,34] and such that the cathode is the one that contributes mostly to the reversible heat generation. As a general rule, for the state of the charge ranges in which the $\partial U/\partial T$ is negative, if the Li atoms intercalate inside the active solid particles then it causes heating and if they de-intercalate from the electrode material it generates cooling. This process is simply opposite when using a range for which the $\partial U/\partial T$ is positive. For the high concentrations of Li atoms inside the $LiCoO_2$ which is the initial state for the battery charging process, the $\partial U/\partial T$ is largely negative which explains the large cooling energy release raised from the reversible heat (as shown in Fig. 3). For the discharging process, the reversible heat presents more complicated pattern as it is initially positive and then turns to be slightly negative and finally becomes largely positive by reaching the saturation limit of the Li atoms concentration inside the $LiCoO_2$ which accordingly releases heating energy originated from the entropy change. Based on our results depicted in Fig. 3, the irreversible heating curves are almost coinciding for charging or discharging process. This highlights the importance of the reversible heating originated from the $LiCoO_2$ cathode in the total heat generation and also explains the underlying mechanism behind the larger positive heat generation during discharging process as compared with the charging loading. It should be however emphasized that the reversible heat is scaled by the applied current in a linear pattern whereas the irreversible heating is similar to the joule heating and scale with a power of two of the current. This implies that the effect of reversible heating is more pronounced for the lower current densities and therefore the difference between the heat generation during charging and discharging is considerably larger for low loading currents. For high C-rates, the effect of  irreversible heating becomes dominant and the net heat generation is always positive.

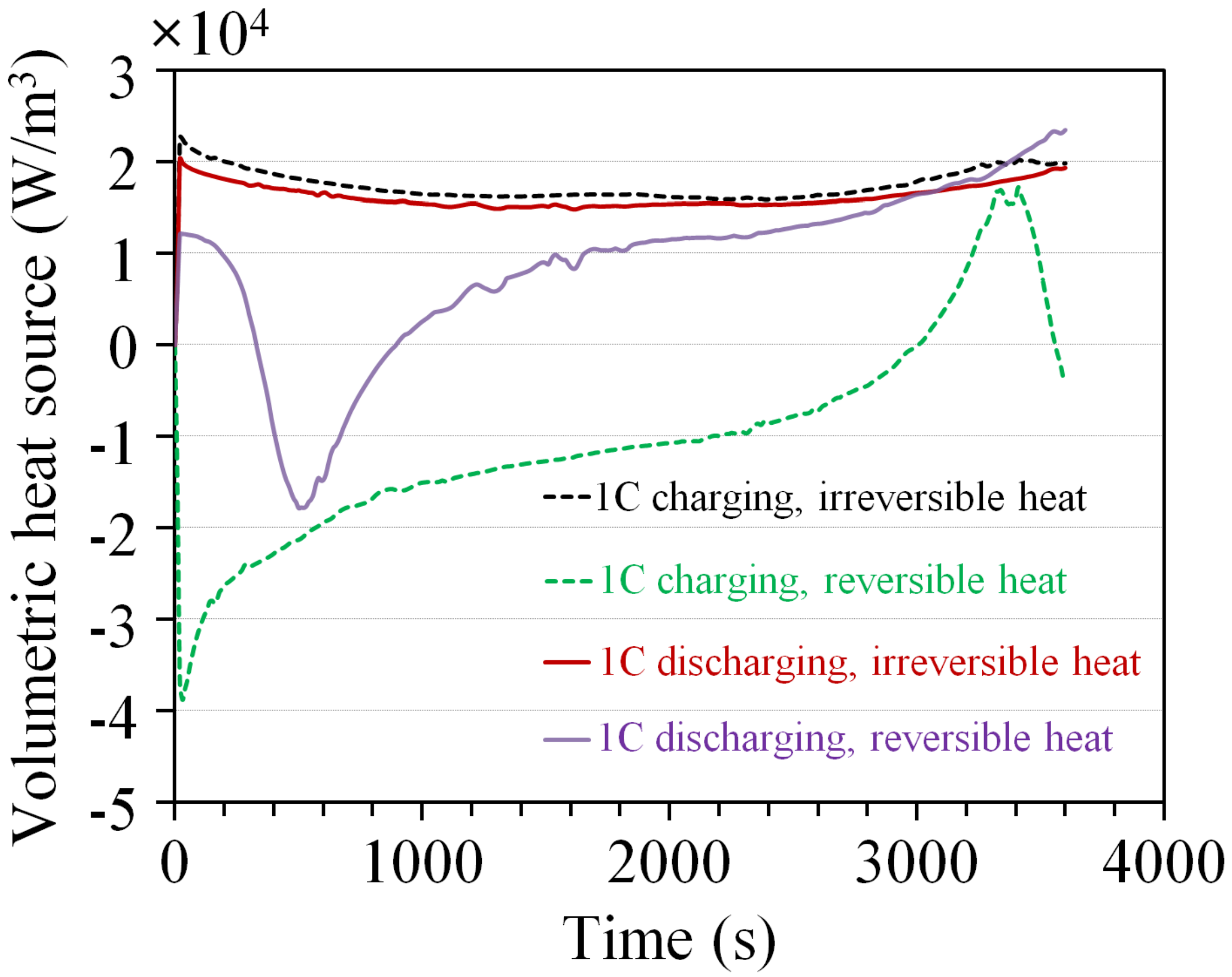

Fig.3, Calculated average volumetric reversible and irreversible heat sources for 1C charging or discharging currents as a function of loading time.

### 2.3 Multiscale modelling and results for thermal properties of paraffin nanocomposites

#### 2.3.1 Molecular dynamics modelling and results for interfacial thermal conductance

In this work, molecular dynamics (MD) simulations were employed to construct atomistic models of multi-layer graphene or h-BN and paraffin interfaces. To do so, first the paraffin structure was constructed and then graphene or h-BN multi-layer films were assembled on the top of the paraffin. Using the developed atomistic composite models, the interfacial thermal conductances between the considered nanomembranes and the paraffin were acquired. All MD calculations in this study were performed using the LAMMPS (*Large-scale Atomic/Molecular Massively Parallel Simulator*) [35] free and open-source package. In the work, paraffin ($C_{30}H_{62}$) wax was simulated as the matrix. Initially a one single molecule with 92 individual atoms was modeled and then it was replicated 729 times to construct a large molecular structure with 67068 atoms. The modeling of the bulk paraffin is very similar to our previous work [13] in which OPLS all-atom force field [36,37] was used for describing atomic interactions of epoxy system. In the OPLS force field, the total energy of the system is the sum of all the individual energies associated with bond, angle, dihedral, and non-bonded interactions. Harmonic approximations were used to simulate bond and angle contributions and OPLS formula was employed to account for dihedral terms. For simplicity, the non-bonded van-der-Waals interaction were introduced by Lennard-Jones (LJ) potential. In this case as a common approach, the equilibrium spacing parameter and potential well depth for each pair were taken to be the arithmetic mean and geometric mean of the native atoms, respectively. A small time step of 0.25 fs was used in our MD simulations. The system was initially equilibrated using Nosé-Hoover thermostat (NVT) method at 300 K. To obtain a homogenous paraffin mixture, the structure was heated to 1000 K and it was kept at this temperature to let paraffin molecules to rearrange their atomic positions. Then the structure was cooled to room temperature using the Nosé-Hoover barostat and thermostat (NPT) method. We note that in this work non-reactive molecular dynamics simulations were conducted, which means that the structural information were fixed during the heating and cooling processes and such that bond creation and rupture were not allowed. After obtaining the homogenous molecular models of paraffin, the graphene or h-BN multi-layer films were assembled on the top of the bulk paraffin.

In this investigation, the optimized Tersoff potentials proposed by Lindsay and Broido for graphene [38] and h-BN [18] were used to introduce the atomic interactions for nanofillers. To the best of our knowledge, Tersoff potentials [39] parameters developed by Lindsay and Broido [18,38] are the most accurate choices to simulate the thermal properties of graphene and h-BN because of the fact that they accurately reproduce the phonon dispersion curves of graphite and bulk h-BN [18,38]. It should be noted that the heat transfer between the fillers and paraffin strongly depends on the assumptions for their corresponding non-bonding interaction. Like our modeling for paraffin atoms, the non-bonded interactions between nanofillers and matrix atoms were simulated using the LJ potential. For boron and nitrogen atoms in h-BN, values of 0.32 nm and 4 meV were used for the equilibrium distance and potential well depth, respectively, as proposed by Lindsay and Broido [18]. On the other side, for carbon atoms in graphene there exist a wide range of values for LJ potentials. However, for the equilibrium distance and potential well depth of carbon atoms in graphene, values of 0.34 nm and 2.4 meV were used, respectively, as suggested by Girifalco *et al.* [40], which has been the most used assumption to the best of our knowledge. The same mixing rule as that used for the paraffin atoms was utilized to introduce LJ interaction parameters between the paraffin and the fillers atoms. Samples of the constructed all-atoms molecular models of multi-layer graphene or h-BN paraffin nanocomposites are illustrated in Fig. 4a and Fig. 4b.

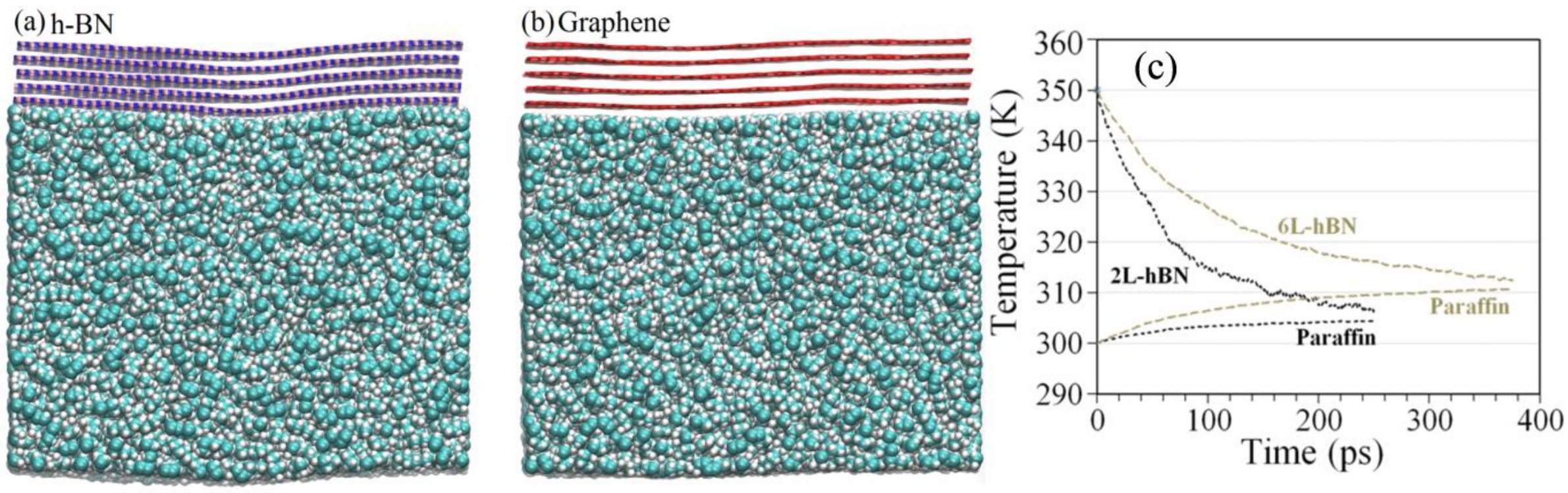

Fig.4, Constructed atomistic models of multi-layer (a) h-BN and (b) graphene paraffin nanocomposites to calculate the interfacial thermal conductance between the paraffin matrix and the nanofillers. (c) Typical averaged temperature profiles for two different nanocomposite structures, two-layer h-BN (2L-hBN)/paraffin and six-layer h-BN (6L-hBN)/paraffin, showing exponential decays as a function of simulation time.

In order to compute the interfacial thermal conductance between the nanofillers and the paraffin atoms, after initial equilibration of the composite structures at the room temperature using the NVT method, an initial temperature difference, $\Delta T(0)$, of 50 K was applied between the fillers and the paraffin. In this case, the paraffin and the nanofillers temperatures were set at 300 K and 350 K, respectively, for 25 ps using the NVT method. To simulate the interfacial heat transfer, the NVT is then switched off and the system was allowed to relax by conducting constant energy (NVE) simulations. The filler and the paraffin atoms temperatures were recorded during the relaxation process as a function of time (t). We note that in order to obtain smooth relations, for every sample six independent simulations were performed with uncorrelated initial atomic velocities and the averaged results were considered for the final calculations (as depicted in Fig. 4c). The temperature difference between the fillers and the matrix, ΔT(t), decays exponentially with a single relaxation time, τ. The interfacial thermal conductance, λ, were obtained using the following relation [41]:

$$\Delta T(t) = \Delta T(0)\, e^{\left[-(1/M_{pa}Cp_{pa} + 1/M_{fil}Cp_{fil})\right]\lambda St} \quad (7)$$

where M and Cp are the mass and the heat capacity, respectively, and S is the contact area of matrix and fillers which is equal to the area of a single graphene or h-BN film in our modeling. In this work the heat capacity of paraffin, graphene, and h-BN at the room temperature were considered to be 2.6 J/gK [42], 0.7 J/gK [43] and 1.6 J/gK [43], respectively.

The predicted interfacial thermal conductance for the graphene/paraffin and the h-BN/paraffin as a function of nanofiller's number of atomic layers are illustrated in Fig. 5. Interestingly, our molecular dynamics results reveal strong dependence of the interfacial conductance on the graphene or the h-BN number of atomic layers for the samples with only few-atomic layers thickness. However, one can observe convergences in the contact thermal conductance as the number of atomic layers reaches four for the both h-BN and graphene multi-layers. Worthy to mention that a similar dependency of the graphene out-of-plane thermal conductivity on the number of atomic layers has been already realized [44]. Our results depicted in Fig. 5 notably reveal that the h-BN films can present much stronger thermal conductance as compared to the graphene films, owing to their strong van-der-Waals interaction with the paraffin atoms as well as their remarkably higher heat capacity. To the best of our knowledge, there exist no direct experimental measurement for the interfacial thermal conductance between the graphene or h-BN with paraffin. Worthy to mention that from experimental point of view, the interfacial thermal conductance between various materials were mostly reported to be below 50 MW/m$^2$K [45].

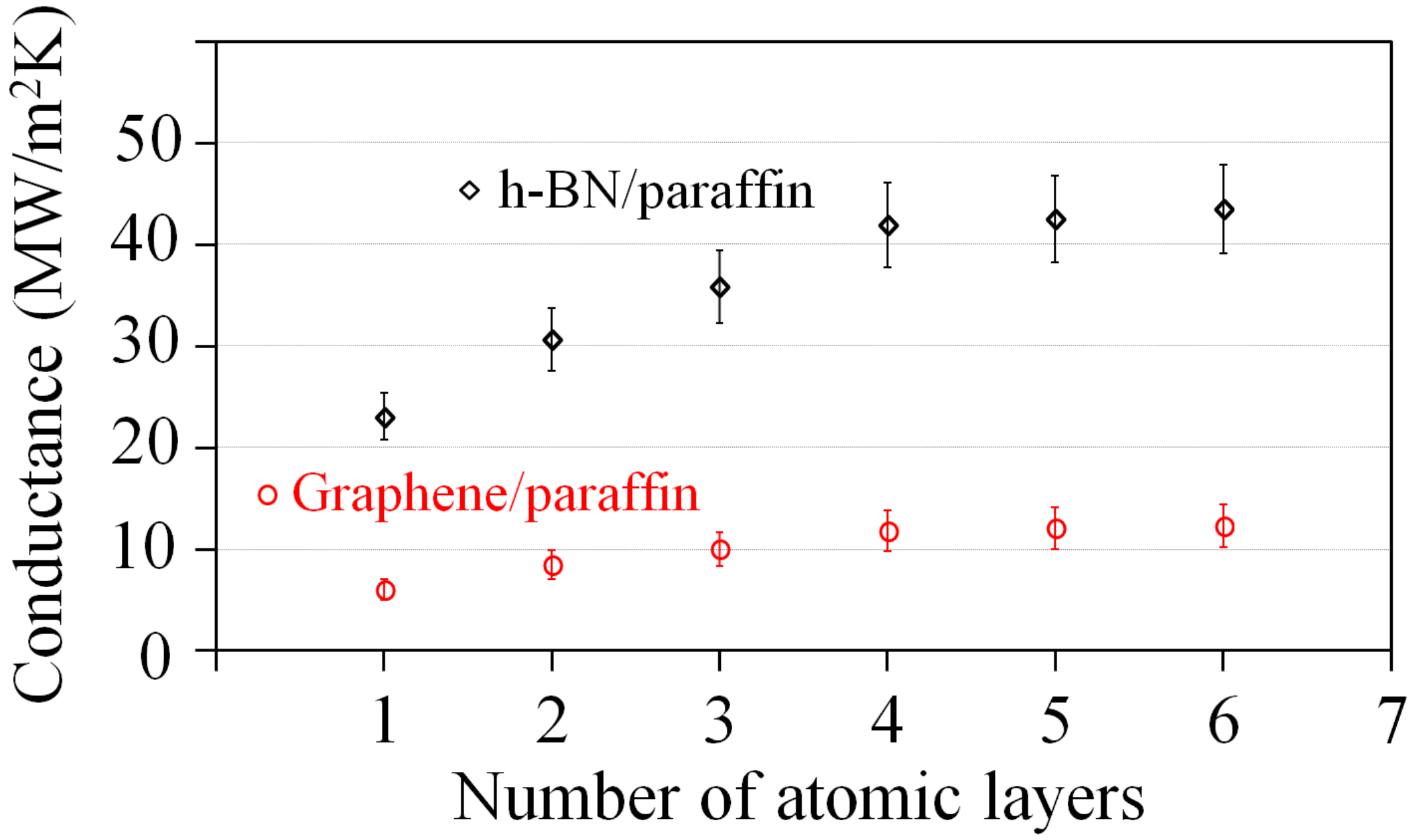

Fig. 5, Molecular dynamics results for the interfacial thermal conductance between the graphene or h-BN and paraffin matrix as a function of fillers number of atomic layers, showing convergence trends for the number of atomic layers higher than four.

### 2.3.2 Finite element modelling and results for the effective thermal conductivity

The effective thermal conductivities of nanocomposites were finally acquired by using the finite element method. Because of high computational costs of the finite element modeling, the simulations of composite structures are limited to modeling of the representative volume elements (RVE) with limited number of fillers. ABAQUS/Standard (Version 6.14) commercial package along with advanced python scripting was used for the modeling of nanocomposites thermal conductivities [13,46,47]. In our modeling, graphene or h-BN nanoplatelets were represented using the disc geometry in which the diameter to thickness ratio stands for the aspect ratio of the fillers. A sample of constructed 3D cubic and periodic

RVE with 4% volume concentration of graphene or h-BN platelets with the aspect ratio of 100 is illustrated in Fig. 6. Because of the computational limitations and the modeling complexities, 300 perfect particles were simulated, which were randomly oriented and distributed in a cube without having intersection which each other [13,46,47]. As explained in-detail in our recent works [13,46,47], in order to calculate the thermal conductivity of the constructed RVEs along a particular direction, a constant heat flux was applied and based on the established temperature difference the thermal conductivity was obtained using the one-dimensional form of the Fourier's law (Fig. 6). Since only limited number of particles were considered to construct the nanocomposite structures with high aspect ratio and high volume concentration for the fillers, the developed RVEs present anisotropic thermal conductivity. In our most recent study [46] it was confirmed that in this case the effective thermal conductivity of the nanocomposite can be well-acquired by averaging the anisotropic thermal conductivities for several RVEs. In this study, for every volume concentration of fillers, four different RVEs were constructed and for each RVE the thermal conductivities along the all three Cartesian directions were computed. The average over the all calculated directional conductivities was reported as the effective thermal conductivity which was found to be acceptably independent of the number of RVEs.

In our finite element modeling of paraffin nanocomposites effective thermal conductivity, a thermal conductivity of 0.23 W/mK [42] was assumed for the paraffin matrix and the thickness of nanofillers were considered to be 10 nm. Worthy to mention that when the number of atomic layers is more than five, the thermal conductivity of the graphene or the h-BN films can be accurately approximated with those for their bulk films [48]. Therefore the thermal conductivity of multi-layer graphene and the h-BN were chosen to be 2000 W/mK [48] and 390 W/mK [18,49], respectively, the same as those for their high quality bulk films. In these RVE modeling, the converged interfacial thermal conductance values obtained by the MD calculations (see Fig. 5) were used to introduce contact conductance for the surface elements between the nanofillers and the paraffin matrix.

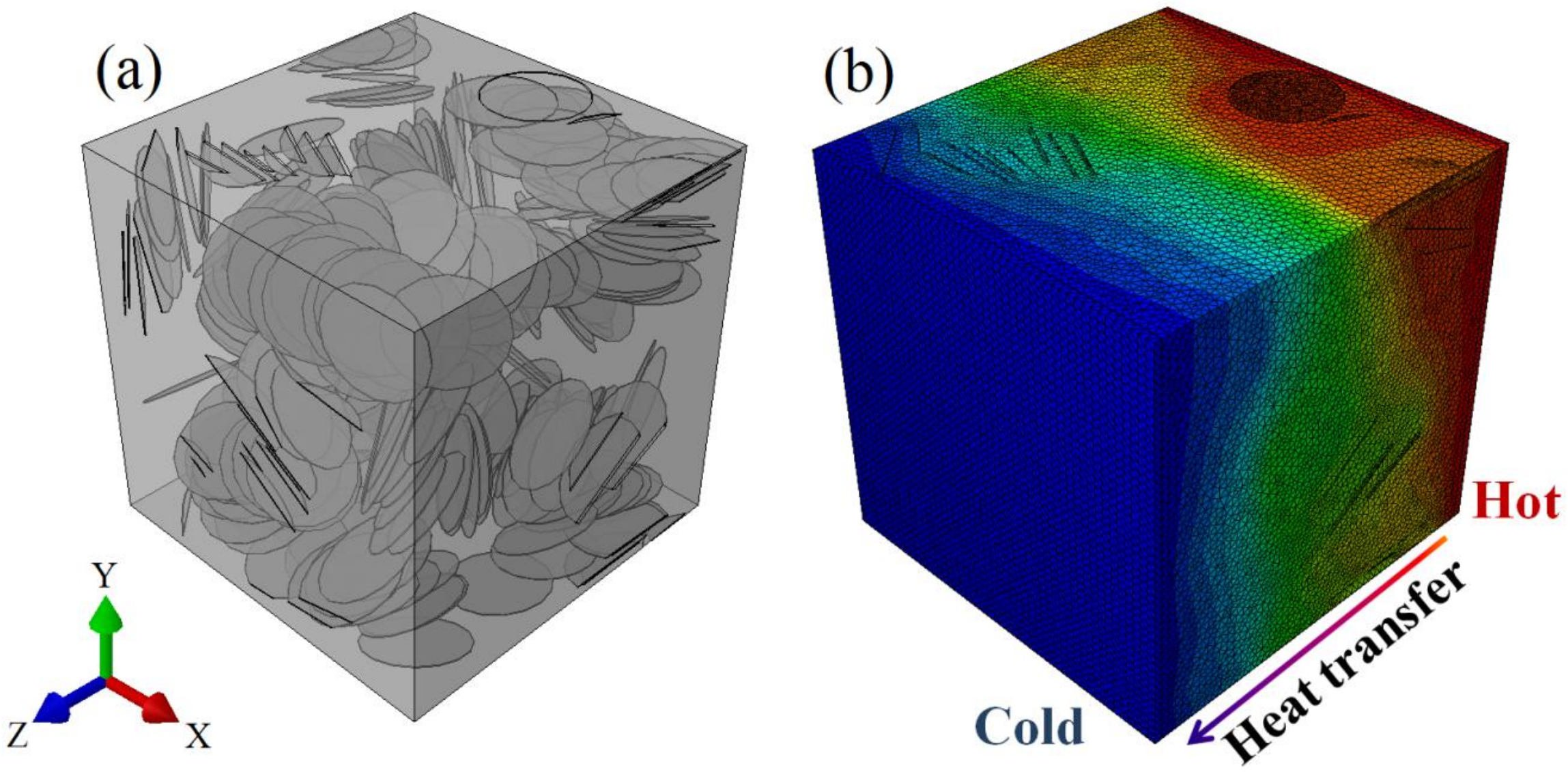

Fig. 6, (a) A sample of constructed 3D and periodic finite element model of paraffin nanocomposite representative volume element (RVE) with 4% concentration of graphene nanofillers with an aspect ratio of 100. To obtain the directional effective thermal conductivity, a constant heat flux was applied along the desired direction and the established temperature difference ($T_{hot}$-$T_{cold}$) was then calculated [46]. (b) Calculated 3D temperature profile for the same sample.

Multiscale simulations were performed for 1% and 4% volume concentrations of nanoplatelets inside the paraffin. The effective thermal conductivities for the paraffin nanocomposites with 1% and 4% volume concentrations of graphene nanofillers were obtained to be 0.35 W/mK and 0.6 W/mK, respectively. From experimental point of view, in agreement with our theoretical results, the thermal conductivity of paraffin nanocomposites reinforced with high concentrations of graphene nanoplatelets was reported to be as high as ~0.36 W/mK [19], ~0.6 W/mK [20], and ~0.7 W/mK [21]. As discussed in our recent study [46], the effective thermal conductivity of a nanocomposite is highly sensitive to the fillers aspect ratio and such that the effective thermal conductivity can reach higher values for the more elongated fillers. The discrepancy between different experimental results can be thus attributed to the various aspect ratios and the thickness of the fillers in different experimental samples. We also emphasis that because of the limitations of the finite element modeling, constructing the RVEs with higher aspect ratios get exponentially expensive from the computational point of view. As elaborately discussed in our recent work [46], there exists currently no modeling approach that can accurately simulate the effective thermal conductivity of nanocomposites with fillers with very high aspect ratios. On the other hand, the effective thermal conductivities for the paraffin reinforced with 1% and 4% volume concentration of the h-BN nanofillers were obtained to be 0.36 W/mK and 0.65 W/mK, respectively. Interestingly, our results suggest that the h-BN films, despite of their several times smaller thermal conductivity, may enhance the thermal conductivity of the paraffin more than the graphene with superior thermal conductivity. Such an unexpected finding is only due to the stronger interfacial thermal conductance in the h-BN/paraffin as compared to that for the graphene/paraffin. As discussed in our recent studies [46,50], the interfacial thermal resistance can dominate the heat transfer at nanoscale. Therefore, only for larger thicknesses of the fillers one can observe the remarkable enhancing effects of graphene nanomembranes because of their exceptionally high thermal conduction properties. Therefore, from the experimental aspects, the incorporation of the fillers with only a few nanometer in size may not directly lead to the enhancement in the thermal conductivity because that may increase the thermal resistance of the composite due to the increase in the interfacial resistances between the matrix and fillers. On the other side, from the modeling point of view, one can match the simulation results with the experimental values only by considering different thicknesses or aspect ratios for the fillers. The results also further confirm that the enhancement of interfacial conductance through chemical modifications and formation of the chemical bonds can be very promising to enhance the thermal conductivity of the nanocomposite structures. Our multiscale study therefore highlights several critical aspects with respect to the heat transfer along the nanocomposite structures.

### 2.3.3 Effective heat capacity of the paraffin nanocomposites

The heat capacity of the composite can be simply obtained by observing the energy balance. This way, the heat capacity of the paraffin composite can be obtained through the following relation:

$$Cp^{eff} = (Cp_{fil}\,\varepsilon_{fil}\rho_{fil} + Cp_{pa}\,\varepsilon_{pa}\rho_{pa})/(\varepsilon_{fil}\rho_{fil} + \varepsilon_{pa}\rho_{pa}) \quad (8)$$

In this way, the heat capacity of the paraffin nanocomposites reinforced with 4% volume concentration ($\varepsilon_{fil}$=0.04) of the graphene or the h-BN nanoplatelets were calculated to be 2.42 J/gK and 2.51 J/gK, respectively.

## 3. Discussions

In this section, the efficiency in the thermal management of the Li-ion batteries [51,52] through using the paraffin composites structures are discussed. To do so, the main factor for the comparison is considered to be the average temperature of the Li-ion battery cells. In this regard, the average temperature of the cells were computed for various charging or discharging loading conditions and for the different battery systems; without phase change material (no PCM), with pure paraffin (PP), and paraffin nanocomposites (PNC). For the paraffin nanocomposites (PNC models), the effective thermal conductivities of composites made from h-BN or graphene nanofillers were predicted to be close. Thus, the results for the 4% volume concentration of h-BN nanomembranes inside the paraffin were considered because of their higher heat capacity.

In Fig. 7, the average temperature of the batteries for 1C discharging or charging currents as a function of loading time are compared. For these cases, the convective heat transfer coefficient was considered to be 10 W/m$^2$K. As it can be observed, the temperature of the batteries in the discharging process are raised considerably higher as compared to the charging process. As discussed for the results depicted in Fig. 3, because of the significance of the entropy changes inside the $LiCoO_2$ cathode, the reversible heat sources for the charging and the discharging are cooling and heating, respectively. Therefore at initial stages of the charging for the 1C charging current, considerable cooling of the battery occurs in which the $LiCoO_2$ cathode is rich of Li atoms and thus the entropy change leads to the overall cooling of the battery. Such an observation is in agreement with the results obtained by the previous studies [27,34]. For the charging, the increase in the temperature of the batteries are generally insignificant as compared to the discharging process. Based on the obtained results therefore the thermal management of the batteries with the $LiCoO_2$ cathode and the graphite anode is pronounced mostly for the discharging process. In this way, the discharging process was considered in the rest of this study discussions.

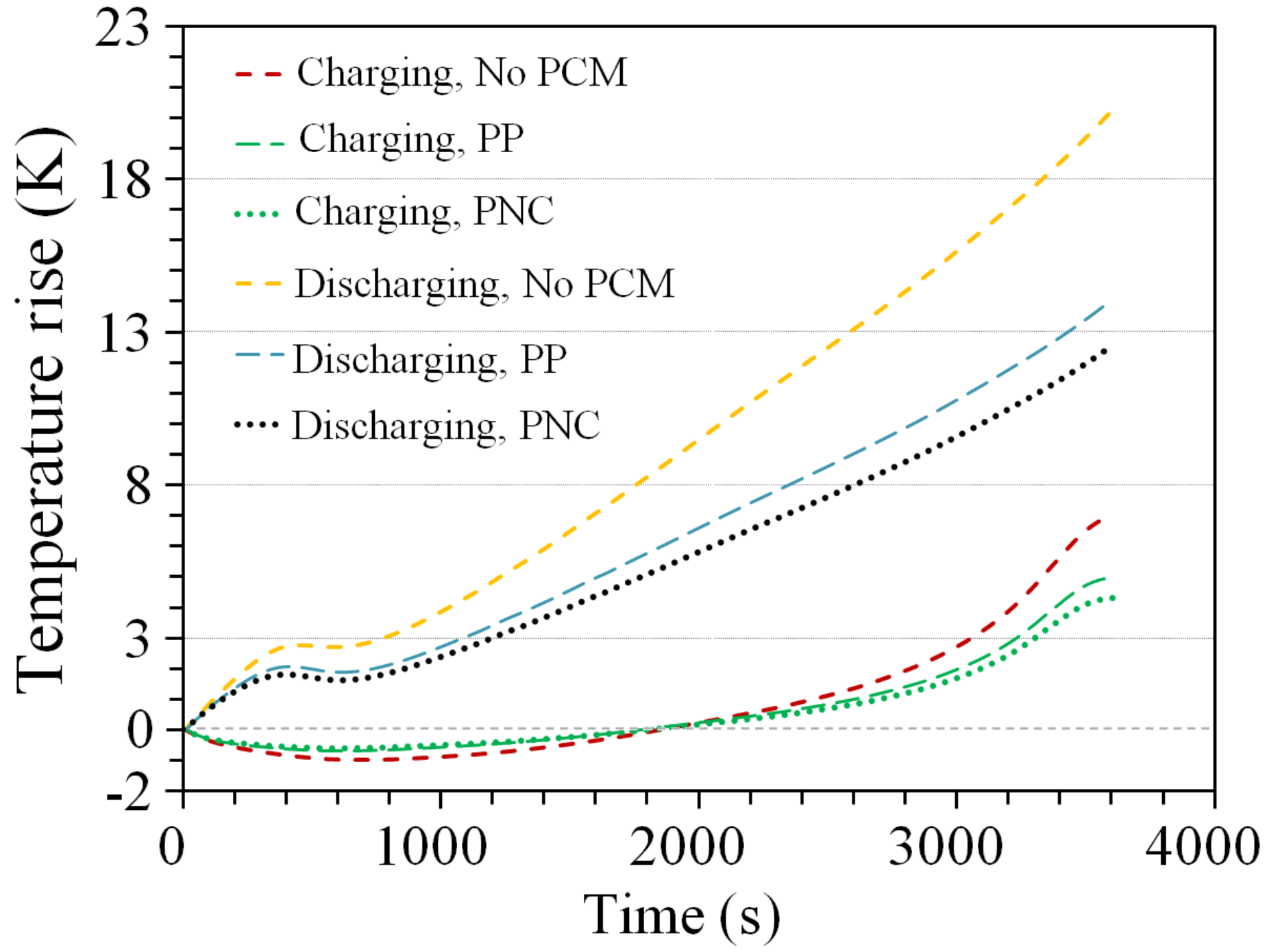

Fig. 7, Calculated average temperature of batteries for 1C discharging or charging currents as a function of loading time. Different battery systems were simulated; without phase change material (No PCM), with pure paraffin (PP) and paraffin nanocomposites (PNC). The convective heat transfer coefficient for these models was considered to be 10 $W/m^2K$.

In Fig. 8, the multiscale results for the average temperature of the batteries for 0.2C, 1C, and 3C discharging currents as a function of loading time for various thermal management methods are compared. As expected, by increasing the discharging current density the temperature rise in the batteries increases significantly which is naturally because of the remarkable increase in the irreversible ohmic heat generation during the battery operation. On the other side, one can observe that in all cases employing the paraffin structures result in distinct improvement in the thermal management. However, our results suggest that the fabrication of the paraffin nanocomposites does not necessarily yield significant improvement in thermal management over the pure paraffin. We remind that our multiscale modelling predictions in accordance with the experimental results [19–21] reveal that the thermal conductivity of the paraffin can be enhanced by several times through addition of the graphene nanoplatelets. However, provided that the addition of nanofillers does not change the chemistry of the material, due to their lower heat capacity as compared to the native paraffin, the heat capacity of the composite declines. Therefore, the addition of highly conductive nanofillers may improve the thermal conductivity of PCM by several times, however, at the same time it may yield undesirable outcomes by suppressing the heat capacity of the PCM. Consequently, the h-BN fillers owing to their higher heat capacity are in principal more promising for the enhancement of the thermal management of the Li-ion batteries in comparison with the graphene nanoplatelets.

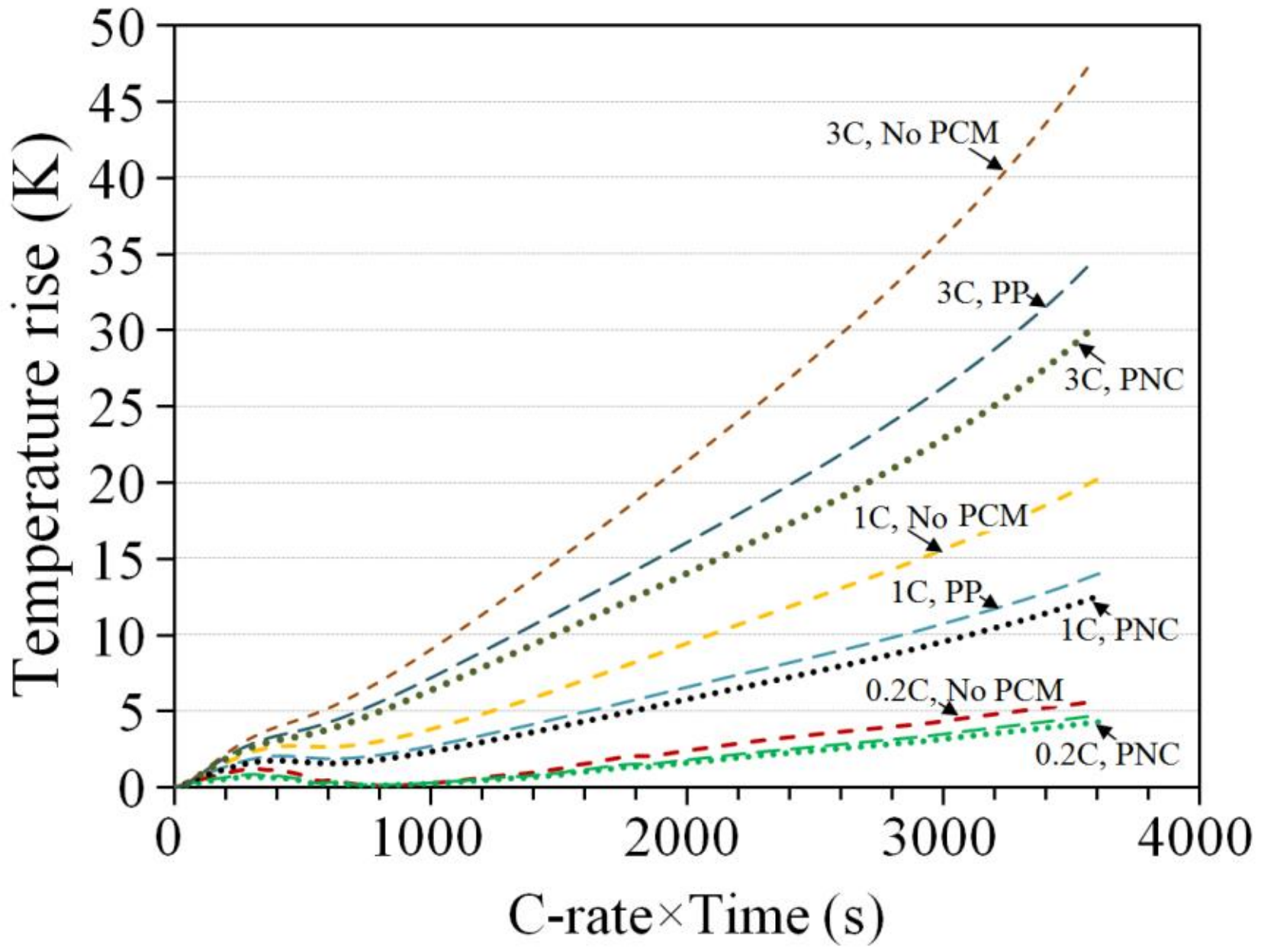

Fig. 8, Multiscale results for the temperature rise of Li-ion batteries for different discharging current densities as a function of time. The convective heat transfer was considered to be 10 $W/m^2K$.

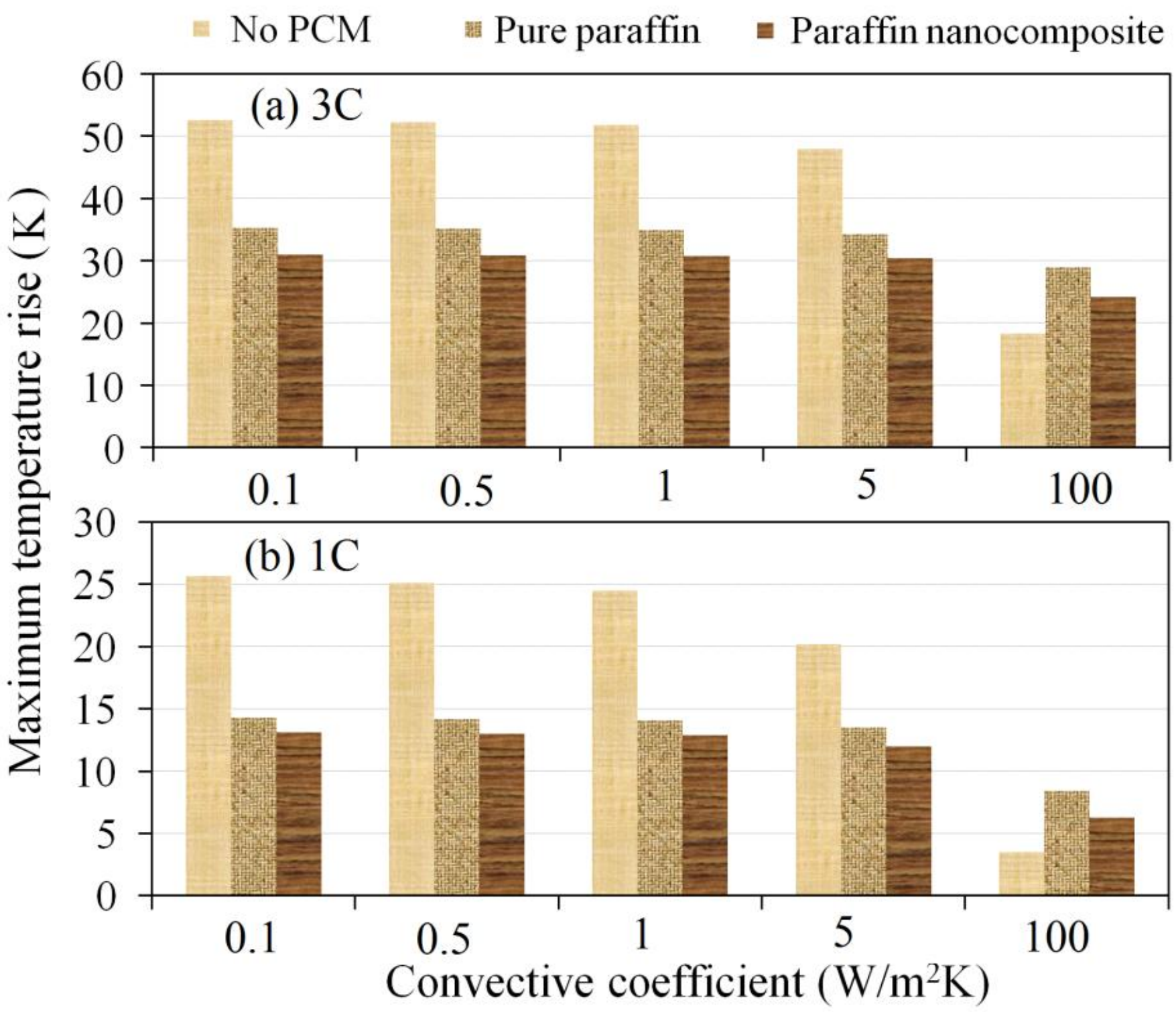

Fig. 9, Modelling results for the effects of different convective heat transfer coefficients between the outer surfaces of the battery system and the ambient air, on the maximum temperature rises of the batteries for two discharging current densities of (a) 3C and (b) 1C.

Another factor that may substantially affect the temperature rise in the Li-ion batteries is the convective heat transfer to the ambient. In Fig. 9, the effects of the various convective heat transfer coefficients between the outer surfaces of the battery system and the ambient air on the maximum temperature rise of the batteries for two different discharging current densities of 1C and 3C are explored. As it is shown, for the 3C discharging current the effect of the convective heat transfer turns to be considerable for the values higher than 1 $W/m^2K$ and before that its effect is convincingly negligible. Interestingly, for the very high convective heat transfer coefficient of 100 $W/m^2K$, the use of PCM leads to an adverse thermal management. In this case the convective heat transfer to the environment is much faster than the heat storage in the PCM and such that the PCM only delays the cooling process. Nonetheless, for the normal conditions, a hybrid system (air-cooling + PCM) provides the more efficient thermal management [53]. Based on our results, the paraffin nanocomposites can reduce the temperature rise by almost 10% as compared to the pure paraffin. Such a small enhancement can be justifiable only for the very precise systems in which the increase in the costs of the PCM structure through the fabrication of the nanocomposites are acceptable.

As discussed earlier, the heat transfer through the nanocomposites depends on the numerous factors such as the quality, the volume concentration, the geometry and the thickness of the fillers, the interfacial thermal conductance between the fillers and the matrix, and the quality as well as the homogeneity of the particles dispersions inside the matrix. From the experimental point of view, because of the statistical nature of the problem, providing the optimum conditions to reach the highest accessible thermal conductivities for the nanocomposites has not been so far achieved and that can well-explain the remarkable variations in the reported experimental results [10,19–21]. Nevertheless, an alternative promising approach to enhance the thermal conductivity of the polymers is to incorporate the networks of highly conductive fillers as proposed by Burger *et al.* [54]. In this case [54], many complexities existing in the nanocomposites are avoided and the experimental results are easier to reproduce. Therefore, in this study, the effects of incorporation of the graphite networks on the thermal management of batteries (see Fig. 10d) are also studied. In accordance with the experimental results, the thermal conductivity and the thickness of the graphitic films were considered to be 200 W/mK and 0.5 mm, respectively. The volume fraction of the graphite films was also modelled to be 4% to match with our nanocomposite samples. In addition, the use of the graphite rings with a distance from the Li-ion batteries with the similar characteristics (see Fig. 10e) are also considered. The effects of the various paraffin composite structures on the maximum temperature rise of the batteries for the various convective heat transfer coefficients and for two different discharging current densities of 1C and 3C are compared in Fig. 10. Our results reveal that the graphite rings present declined performance in the thermal management even in comparison with the pure paraffin. In this case, the rings focus the heat toward the batteries instead of transferring them away and such that they are not recommended for the thermal management of batteries. Interestingly, the graphite networks in comparison with other considered systems yield the best performance in the thermal management of the battery system. Unlike the statistical nature of the nanocomposite structures, the graphite networks can be arranged in an optimized pattern to more efficiently engineer the thermal management of the batteries. However, more detailed investigations should be conducted in this direction which mainly deal with the optimization procedures in heat transfer problems. Nevertheless, our study highlights that the graphite or the h-BN flakes networks paraffin composites can be considered as one of the most promising approaches for the advanced thermal management of rechargeable-ion batteries in response to the overheating concerns.

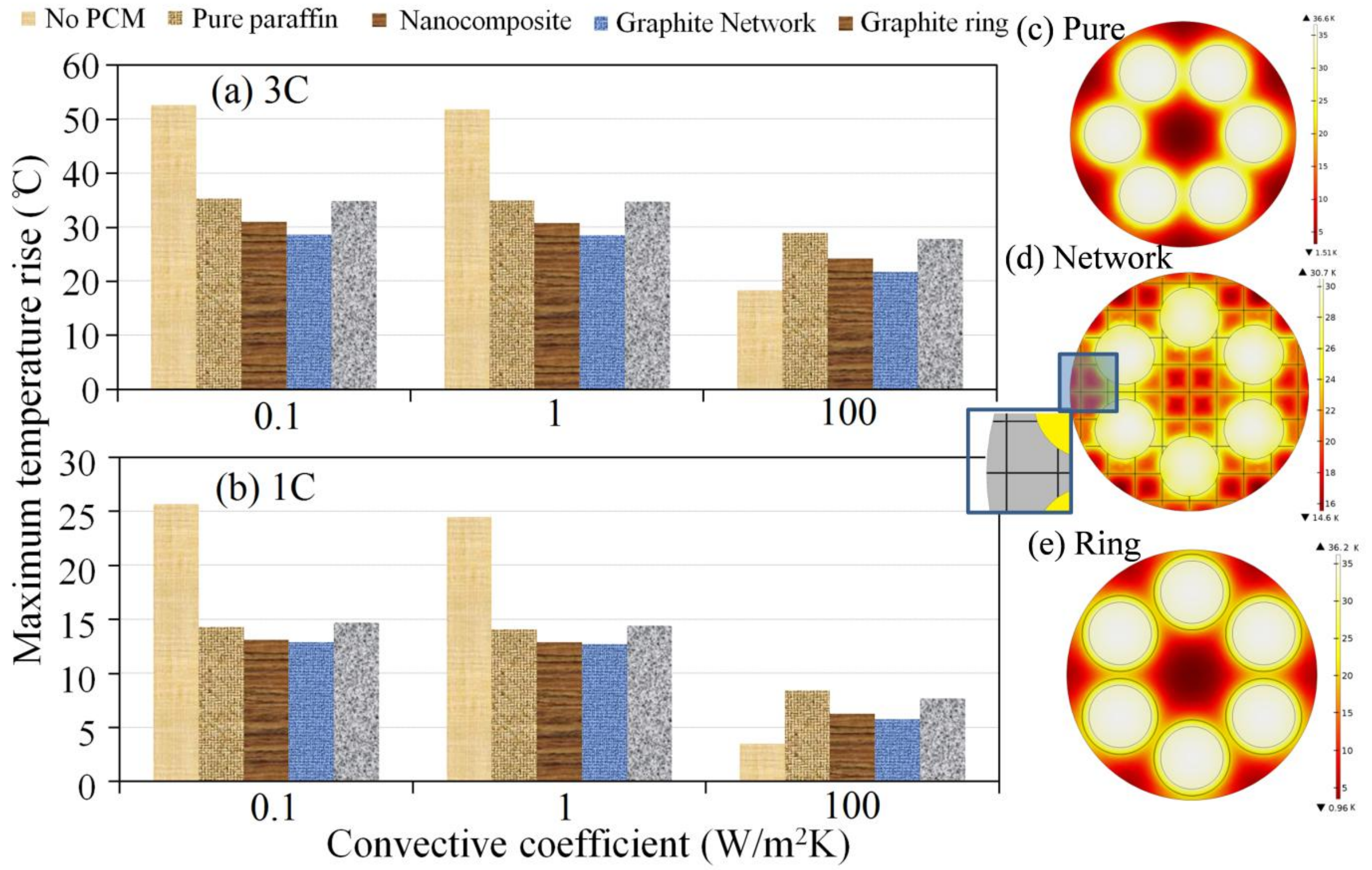

Fig. 10, Effects of various paraffin composite structures on the maximum temperature rises of the batteries for two different discharging current densities of (a) 3C and (b) 1C and for various convective heat transfer coefficients. Top views of the battery systems temperature profile at the maximum temperature rise for (c) pure paraffin, (d) graphite network (the inset shows the zoom of the structure) and (e) graphite ring.

## 4. Summary

Extensive multiscale modelling were conducted to explore the efficiency in the enhancement of Li-ion batteries thermal management through employing the paraffin composite structures. First, the electrochemical processes of a Li-ion battery was simulated based on the Newman's pseudo 2D electrochemical model, aiming to acquire the total average volumetric heat generation functions for different charging or discharging current densities. The results illustrate the importance of the reversible heat source originated from the $LiCoO_2$ cathode in the total heat generation and also suggest the larger heat generation during discharging over the charging process.

The effective thermal conductivity of paraffin nanocomposites reinforced with graphene or h-BN nanomembranes were acquired by developing combined MD/FE multiscale modelling. Multiscale results highlight several important aspects with regard to the heat transfer through nanocomposite structures:

- Modeling results reveal initial strong dependence of the interfacial conductance between the paraffin polymer and graphene or h-BN nanomembranes on the fillers number of atomic layers. However, it was found that the interfacial thermal conductance converges for the nanofillers with atomic layers more than four.
- Atomistic results suggest that h-BN nanomembranes may present much stronger thermal conductance with paraffin as compared to the graphene nanofilms.
- Multiscale simulations illustrate the dominance of interfacial thermal conductance for the heat transfer in nanocomposites. Interestingly, it was predicted that h-BN nanofilms owing to their stronger interfacial thermal conductance with paraffin atoms may enhance the thermal conductivity of the paraffin more than the graphene nanosheets.
- Thermal conductivity and heat capacity of polymers filled with h-BN nanofilms can be potentially higher than those for the same polymers reinforced with graphene films with similar geometrical characteristics. By taking into consideration that the h-BN/polymer nanocomposites remain electrically insulators, they might be considered as promising building blocks in response to overheating concerns, particularly in nanoelectronics.

Finally the efficiency in the thermal management of a battery system was investigated through considering: no phase change material, pure paraffin, paraffin nanocomposites, and graphite network paraffin composite. In this case, the acquired heat generation functions based on the electrochemistry solutions were used to represent the heat sources during the battery operation. It was shown that by employing the pure paraffin, remarkable improvement in the thermal management can be accomplished. Nevertheless, it was concluded that the fabrication of the paraffin nanocomposites does not yield significant improvement in the Li-ion batteries thermal management over the pure paraffin and such that their application can be recommendable only for limited cases. Interestingly, among the all studied cases our study highlights that the graphite networks paraffin composites with optimized configurations can be considered as one of the most appealing solutions for the advanced thermal management of rechargeable-ion batteries.

The insights provided by this investigation can be useful to guide the both experimental and theoretical studies for the application of polymer composite structures in response to the thermal management concerns in various systems such as the batteries and nanoelectronics.

**Acknowledgment**

BM, FM and TR greatly acknowledge the financial support by European Research Council for COMBAT project (Grant number 615132). The authors specially thank Andreas Hess, Quirina Roode-Gutzmer and Manfered Bobeth at TU-Dresden for fruitful discussions. HY and GC acknowledge the support by the German Research Foundation (DFG) within the Cluster of Excellence "Center for Advancing Electronics Dresden" (cfAED), the Center for Information Services and High Performance Computing (ZIH) at TU-Dresden for computational resources and the support by Dresden Center for Computational Materials Science (DCCMS).

## Nomenclature

$a_s$ active surface area per electrode unit volume ($m^2/m^3$)
$c$ concentration of Li in a phase (mol/ $m^3$)
$M$ mass of a phase (gr/mol)
$Cp$ heat capacity of a phase (J/ gK)
$V$ cell voltage (V)
$D$ diffusion coefficient of lithium species ($m^2/s^1$)
$F$ Faraday's constant, 96,487 C /$mol^1$
$h$ convective heat transfer coefficient (W/$m^2$ $K^1$)
$i_0$ exchange current density of an electrode reaction ($A/m^2$)
$I$ applied current (A)
$S$ contact surface ($m^2$)
$SOC$ state of the charge
$j^{Li}$ reaction current density (A /$m^3$)
$K$ rate constant for electrochemical reaction of electrode (A $m^{2.5}/mol^{1.5}$)
$L$ domain's length (µm)
$Q_j$ volumetric ionic heat generation from the motion of Li through the solid and electrolyte phases (W/$m^3$)
$Q_r$ volumetric heat generation from the reaction current and over-potentials (W/$m^3$)
$Q_{rev}$ volumetric reversible heat generation from entropy changes in the solid particles (W/$m^3$)
$r$ radial coordinate along active material particle (cm)
$R$ universal gas constant, 8.3143 J/mol K
$R_s$ average radius of solid particles (m)
$t$ time (s)
$t^0_+$ transference number of Li ion inside electrolyte
$T$ absolute temperature (K)
U open-circuit voltage of an electrode reaction (V)
$\Delta T$ temperature difference between the fillers and paraffin matrix (K)

***Greek symbols***

$\varepsilon$ volume fraction of a phase
$\eta$ surface over-potential of an electrode reaction (V)
$\kappa$ ionic conductivity of electrolyte (S/m)
$\kappa_D$ diffusional conductivity of a species (A /m)
$\sigma$ conductivity of solid particles (S /m)
$\varphi$ electrical potential in a phase (V)
$\tau$ relaxation time in the temperature difference (s)
$\lambda$ interfacial thermal conductance between fillers and matrix (s)
$\rho$ density ($kg/m^3$)

***Subscripts***

i anode, separator, or cathode region
e electrolyte phase
s solid phase
b binder for solid particles in the electrode
max maximum concentration
ref with respect to a reference state
s,surf Li concentration on the surface of particles
sep separator region
avg average
n negative electrode region
p positive electrode region
pa paraffin matrix
fil graphene or h-BN fillers

***Superscripts***

eff effective property of a medium
Li lithium species
brug Bruggeman effective medium factor

Table 1, Governing equations and boundary conditions according to the Newman's pseudo-2D model [24]. The effective properties [33] were obtained mainly by using the Bruggeman's relation.

| **Electrochemical equations** | | **Boundary conditions** |
|---|---|---|
| Mass balance, electrolyte phase | $\frac{\partial(\varepsilon_e c_e)}{\partial t} = \frac{\partial}{\partial x}\left(D_e^{\text{eff}} \frac{\partial}{\partial x} c_e\right) + \frac{1-t_+^0}{F} j^{Li}$ | $\frac{\partial c_e}{\partial x}\vert_{x=0} = \frac{\partial c_e}{\partial x}\vert_{x=L_n+L_{sep}+L_p} = 0$ |
| Mass balance, solid phase | $\frac{\partial c_s}{\partial t} = \frac{D_s}{r^2}\frac{\partial}{\partial r}\left(r^2 \frac{\partial c_s}{\partial r}\right)$ | $\frac{\partial c_s}{\partial r}\vert_{r=0} = 0, -D_s \frac{\partial c_s}{\partial r}\vert_{r=R_s} = \frac{j^{Li}}{a_s F}$ |
| Electrical potential, electrolyte phase | $\frac{\partial}{\partial x}\left(\kappa^{\text{eff}} \frac{\partial}{\partial x} \varphi_e\right) + \frac{\partial}{\partial x}\left(\kappa_D^{\text{eff}} \frac{\partial}{\partial x} \ln c_e\right) + j^{Li} = 0$ | $\frac{\partial \varphi_e}{\partial x}\vert_{x=0} = \frac{\partial \varphi_e}{\partial x}\vert_{x=L_n+L_{sep}+L_p} = 0$ |
| Electrical potential, solid phase | $\frac{\partial}{\partial x}\left(\sigma^{\text{eff}} \frac{\partial}{\partial x} \varphi_s\right) = j^{Li}$ | $-\sigma_n^{\text{eff}} \frac{\partial \varphi_s}{\partial x}\vert_{x=0} = \sigma_p^{\text{eff}} \frac{\partial \varphi_e}{\partial x}\vert_{x=L_n+L_{sep}+L_p} = \frac{I}{A}$<br>$\frac{\partial \varphi_s}{\partial x}\vert_{x=L_n} = \frac{\partial \varphi_s}{\partial x}\vert_{x=L_n+L_s} = 0$ |
| **Effective properties** | | |
| Electrolyte ionic diffusivity | | $D_e^{\text{eff}} = D_e \varepsilon_e^{brug}$ |
| Electrolyte ionic conductivity | | $\kappa^{\text{eff}} = \kappa \varepsilon_e^{brug}$ |
| Electrolyte ionic diffusion conductivity | | $\kappa_{\text{D}}^{\text{eff}} = \frac{2RT\kappa^{\text{eff}}}{F}(1 - t_+^0)$ |
| Solid phase electronic conductivity | | $\sigma^{\text{eff}} = \varepsilon_s \sigma$ |

Table 2, Parameters used for the Li-ion battery cell electrochemical modelling.

| **Parameter** | **Description** | **Negative electrode** | **Separator** | **Positive electrode** |
|---|---|---|---|---|
| $t^0$ | Transference number of electrolyte [27] | 0.435 | 0.435 | 0.435 |
| $L$ | Length of electrode (μm) [27,55] | 73.5 | 25 | 70 |
| $r$ | Solid particle radius (μm) [27,55] | 12.5 | — | 8.5 |
| $\varepsilon_s$ | Solid phase volume fraction[55] | 0.50 | — | 0.55 |
| $\varepsilon_e$ | Electrolyte phase volume fraction[55] | 0.44 | 0.45 | 0.30 |
| $\varepsilon_b$ | Binder volume fraction [27] | 0.0566 | — | 0.15 |
| $D_s$ | Solid phase diffusion coefficeint ($m^2$/s) [56] | $3.89\times10^{-14}$ | — | $10\times10^{-14}$ |
| $P$ | Active material density (kg/$m^3$) [56] | 2292 | — | 5032 |
| $C_{s,max}$ | Maximum concentration in solid phase (mol/$m^3$) [27,55] | 31 858 | — | 49 943 |
| $SOC_{,0}$ | Initial state of charge in Charge/Discharge (%) | 0.02/0.98 | — | 0.97/0.45 |
| $C_{e,0}$ | Initial electrolyte concentration (mol/$m^3$) [27] | 1000 | 1000 | 1000 |
| *brug* | Bruggeman coefficient for tortuosity [27] | 4.1 | 2.3 | 1.5 |
| $K_i$ | Reaction rate coefficient (A $m^{2.5}$/$mol^{1.5}$) [27] | $1.764\times10^{-11}$ | — | $6.67\times10^{-11}$ |
| $\sigma$ | Solid phase conductivity (S/m) [27,55] | 100 | — | 10 |